\documentstyle[twocolumn,prl,aps,graphicx,floats]{revtex}
\begin{document}
\draft
\twocolumn[\hsize\textwidth\columnwidth\hsize\csname@twocolumnfalse\endcsname

\title{Origin of the anomaly in diffuse scattering from 
disordered Pt-V alloys}
\author{Igor Tsatskis\cite{pers}}
\address{Department of Earth Sciences, University of Cambridge,
Downing Street, Cambridge CB2 3EQ, United Kingdom}
\maketitle

\begin{abstract}
An explanation of the anomalous concentration dependence of diffuse 
scattering from the Pt-V alloy system (splitting of the (100) short-range 
order intensity peak with increasing Pt content) is proposed. The effect 
is attributed to the competition between the interaction and self-energy 
curvatures. A similar temperature behaviour is predicted.
\end{abstract}

]

In this Letter we put forth an explanation of recently 
published~\cite{lebolloch} unusual experimental and Monte Carlo 
(MC) results for diffuse scattering from disordered Pt-V alloys. 
The short range order (SRO) diffuse intensity (hereafter referred 
to simply as the intensity) for two alloys, Pt$_{3}$V at 1393~K and 
Pt$_{8}$V at 1224~K, was obtained in Ref.~\cite{lebolloch} from in 
situ neutron scattering data by separating different contributions 
to the total intensity, and a qualitative difference between the 
two intensity distributions was found. The Pt$_{3}$V intensity showed 
a simple peak at the (100) position, while in the case of the 
Pt$_{8}$V alloy this peak was split along the ($h$00) line. The 
Pt$_{8}$V intensity thus exhibited a saddle point, rather than a 
maximum, at the (100) position. On the other hand, the inverse Monte 
Carlo (IMC) interactions determined from the SRO results were nearly 
the same for both alloys (pairwise interactions were assumed), 
indicating that the whole effect might be of statistical-mechanical 
origin. To verify this conjecture, the set of the IMC interactions 
found for the Pt$_{3}$V alloy was used to generate the MC intensity 
for the Pt$_{8}$V composition at 410 K (50 K above the calculated 
MC transition temperature), and the splitting of the (100) peak was 
indeed observed. At the same time, the MC intensity for the Pt$_{3}$V 
alloy was in good agreement with its experimental counterpart, 
showing no splitting. The conclusion is, therefore, that the same 
interaction can lead to intensity peaks at different positions, 
depending on concentration (and, in fact, on temperature, since the 
measurements and MC simulations for the two alloys were carried 
out at different temperatures). This conclusion is in clear 
contradiction with the standard mean-field arguments, according 
to which peaks of the intensity should always correspond 
to minima of the pair interaction in the reciprocal space.

Thus, the question arises about the origin of the observed anomaly. 
It is shown below that such behaviour is the result of the competition 
between the reciprocal-space curvatures of the interaction and 
self-energy terms entering the formula for the intensity 
on an equal footing. The splitting develops when the self-energy 
curvature exceeds the curvature of the interaction term. In the 
mean-field approach the wavevector dependence of the self-energy 
is ignored, hence the inability to account for effects similar to 
observed in Ref.~\cite{lebolloch}. Indeed, the formally exact expression 
for the SRO diffuse intensity is~\cite{tokar1}
\begin{equation}
I({\bf k}) = \frac{1}{ c(1-c) \left[ - \Sigma({\bf k}) + 
2 \beta V({\bf k}) \right] } \ . \label{1}
\end{equation}
Here ${\bf k}$ is the wavevector, $I({\bf k})$ the intensity in Laue 
units, $c$ the concentration, $\beta=1/T$, $T$ the temperature in 
energy units, $V({\bf k})$ the Fourier transform of the pair ordering 
potential $V_{ij}$,
\begin{equation}
V_{ij}=\frac{1}{2} (V^{AA}_{ij}+V^{BB}_{ij}) - V^{AB}_{ij} \ , \label{2}
\end{equation}
and the potential $V^{\alpha \beta}_{ij}$ corresponds to the interaction 
between an atom of type $\alpha$ at site $i$ and an atom of type $\beta$ 
at site $j$. Finally, $\Sigma({\bf k})$ is the self-energy of the pair 
correlation function (PCF); the PCF in the ${\bf k}$-space representation 
is proportional to $I({\bf k})$. Apart from the wavevector, $\Sigma$ also 
depends on the concentration and the temperature. Eq.~(\ref{1}) is one of 
the possible forms of the Dyson equation~\cite{izyumov} which is 
satisfied by the PCF; this issue is discussed in considerable detail 
elsewhere~\cite{tsatskis}. In the mean-field Krivoglaz-Clapp-Moss (KCM) 
approximation~\cite{krivoglaz},
\begin{equation}
I^{KCM}({\bf k}) = \frac{1}{1 + 2 c(1-c) \beta V({\bf k})} \ , \label{3}
\end{equation}
however, $\Sigma$ is a function of concentration only,
\begin{equation}
\Sigma^{KCM} = - \frac{1}{c(1-c)} \ . \label{4}
\end{equation}
Another, much better approximation for $I({\bf k})$ is the spherical 
model (SM)~\cite{brout} (also known as the Onsager cavity field 
theory~\cite{onsager}), 
\begin{equation}
I^{SM}({\bf k}) = \frac{1}{ c(1-c) \left[ - \Sigma^{SM} + 
2 \beta V({\bf k}) \right] } \ ,  \label{15}
\end{equation}
where $\Sigma^{SM}$ is a number determined from the sum rule
\begin{equation}
\alpha_{000} = \frac{1}{\Omega} \int d {\bf k} \, 
I({\bf k}) = 1 \ . \label{16}
\end{equation}
Here $\alpha_{lmn}$ is the SRO parameter for the 
coordination shell $lmn$, $I({\bf k})$ is the Fourier transform of 
$\alpha_{lmn}$, and the integration is carried out over the Brillouin 
zone of volume $\Omega$. As follows from Eqs.~(\ref{15}) and (\ref{16}),
the SM also takes into account the temperature dependence of the 
self-energy. Nevertheless, in the SM the self-energy is still 
wavevector-independent.

\begin{table}
\caption{Nine IMC pair interactions $V_{lmn}$ for the Pt$_{3}$V alloy 
and first nine SRO parameters $\alpha_{lmn}$ for the Pt$_{3}$V and 
Pt$_{8}$V alloys obtained in Ref.~\protect\cite{lebolloch} from the 
experimental diffuse intensities~\protect\cite{lebolloch1}. Note 
the factor of 2 difference between the definitions of $V_{lmn}$ in 
Ref.~\protect\cite{lebolloch} and the present work.}
\label{t1}
\begin{tabular}{cddd}
$lmn$ & $V_{lmn}$, Pt$_{3}$V & $\alpha_{lmn}$, Pt$_{3}$V & 
$\alpha_{lmn}$, Pt$_{8}$V \\
\tableline
110 &  89.0  & -0.1619 & -0.1234 \\
200 & -14.0  &  0.2144 &  0.1505 \\
211 &  12.6  & -0.0031 & -0.0214 \\
220 &  11.2  &  0.0700 &  0.0215 \\
310 &   9.8  & -0.0589 & -0.0049 \\
222 &   0.38 &  0.0141 &  0.0121 \\
321 &  -1.76 & -0.0065 & -0.0020 \\
400 &   4.6  &  0.0423 &  0.0136 \\
330 &  -8.0  & -0.0029 &  0.0314 \\
\end{tabular}
\end{table}

Meanwhile, in our case the wavevector dependence of the self-energy 
is crucial. Let us consider the behaviour of all ${\bf k}$-dependent 
quantities in Eq.~(\ref{1}) as functions of the deviation $k$ of 
the wavevector from the (100) position along the ($h$00) line. 
We assume that $V(k)$, similarly to the case of the Pt$_{3}$V 
alloy, has a simple minimum at $k=0$ (see Fig.~\ref{f3}b below). 
Mathematically, the presence or absence of the splitting of the (100) 
intensity peak is controlled by the sign of the second derivative 
$I^{\prime \prime} = (\partial^{2} I / \partial k^{2})_{k=0}$. 
From Eq.~(\ref{1}) it follows that
\begin{equation}
I^{\prime \prime} = c(1-c) I^{2}(0) \left( \Sigma^{\prime \prime} 
- 2 \beta V^{\prime \prime} \right) \ . \label{5}
\end{equation}
Eq.~(\ref{5}) shows that the self-energy curvature 
$\Sigma^{\prime \prime}$ can dominate if 
the (100) minimum of $V(k)$ is shallow ($V^{\prime \prime}$ is small). 
In particular, it is $\Sigma^{\prime \prime}$ that controls the 
fine structure (single- vs. double-peak) of the $k=0$ maximum of 
$I(k)$ in the limiting case of vanishing $V^{\prime \prime}$. Thus, 
according to Eq.~(\ref{5}) the splitting occurs when $\Sigma(k)$ 
is a convex-down function and its (100) curvature is greater than 
that of the interaction term $2 \beta V(k)$. In contrast to this, in 
the KCM and SM approximations the self-energy is ${\bf k}$-independent 
(i.e. flat), and the peak splitting occurs only when the interaction 
has a double minimum.

\begin{figure}
\begin{center}
\includegraphics[angle=-90]{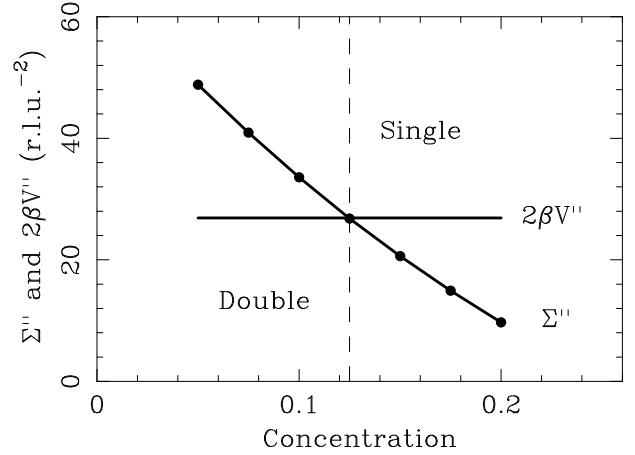}
\end{center}
\caption{Derivatives $\Sigma^{\prime \prime}$ and 
$2 \beta V^{\prime \prime}$ (Eq.~(\ref{5})) as functions of 
concentration at $T=1224$ K. The type of the (100) intensity 
maximum (single or double peak) is indicated.}
\label{f1}
\end{figure}

Our next step is to make some estimates using available experimental 
and IMC data. Listed in Table~\ref{t1} are the IMC interaction parameters 
$V_{lmn}$ for the Pt$_{3}$V alloy used in Ref.~\cite{lebolloch} in the 
direct MC simulations for both Pt$_{3}$V and Pt$_{8}$V compositions, and 
the SRO parameters $\alpha_{lmn}$ for the two alloys~\cite{lebolloch1}. 
These data can be used to estimate the curvatures of the interaction and 
the self-energy terms in Eq.~(\ref{1}). For an arbitrary matrix $f_{ij}$ 
defined on the FCC lattice, the second derivative $f^{\prime \prime}$ of 
its Fourier transform $f({\bf k})$ is 
\begin{eqnarray}
f^{\prime \prime} & = & 8 \pi^{2} \left( f_{110} - f_{200} 
- 2 f_{211} - 4 f_{220} + 10 f_{310} \right. \nonumber \\
& & \left. - 4 f_{222} + 12 f_{321} - 4 f_{400} 
+ 9 f_{330} \right) \ , \label{6}
\end{eqnarray}
when, as in Table~\ref{t1}, first nine coordination shells are 
taken into account; the factor $\pi^{2}$ reflects the fact that 
the wavevector is measured in reciprocal lattice units (r.l.u.). 
Using Eq.~(\ref{6}), the curvature $V^{\prime \prime}$ of the Pt$_{3}$V 
interaction can be easily calculated, and the result at 1393 K is 
$2 \beta V^{\prime \prime}=23.6$ r.l.u.$^{-2}$. To calculate the curvature 
of the self-energy, we use the expansion of its off-diagonal (in the 
direct-space representation) part in powers of the SRO parameters 
(for short, the $\alpha$-expansion, or AE), of which two first non-zero 
orders are available~\cite{tokar1},
\begin{mathletters}
\label{9}
\begin{eqnarray}
\Sigma_{lmn} & = & a \alpha_{lmn}^{2} + b \alpha_{lmn}^{3} + 
O(\alpha^{4}) \ , \ \ \ lmn \neq 000 \ , \label{9a} \\
a & = & \frac{(1-2c)^{2}}{2[c(1-c)]^{2}} \ , \label{9b} \\
b & = & \frac{[1-6c(1-c)]^{2}-3(1-2c)^{4}}{6[c(1-c)]^{3}} \ . \label{9c}
\end{eqnarray}
\end{mathletters}
Eqs.~(\ref{9}) were derived in the context of the 
$\gamma$-expansion method (GEM)~\cite{tokar1,tokar2}. Using 
Eqs.~(\ref{6}), (\ref{9}) and $\alpha_{lmn}$ from Table~\ref{t1}, 
we get $\Sigma^{\prime \prime}(\mbox{Pt}_{3}\mbox{V})=2.7$ 
r.l.u.$^{-2}$ and $\Sigma^{\prime \prime}(\mbox{Pt}_{8}\mbox{V})=58.5$ 
r.l.u.$^{-2}$. It is immediately seen that
\begin{equation}
\Sigma^{\prime \prime} (\mbox{Pt}_{3}\mbox{V}) < 
2 \beta V^{\prime \prime} (\mbox{Pt}_{3}\mbox{V}) < 
\Sigma^{\prime \prime} (\mbox{Pt}_{8}\mbox{V}) \ , \label{8}
\end{equation}
with a considerable difference between each two values. The fact 
that SRO for the two alloys was measured at different temperatures is 
not very important here; $2 \beta V^{\prime \prime}=26.9$ r.l.u.$^{-2}$ 
at 1224 K. Eq.~(\ref{8}) suggests that the set $\{ V_{lmn} \}$ from 
Table~\ref{t1} leads to the splitting of the (100) intensity peak for the 
Pt$_{8}$V composition at 1224 K, whereas in the case of the Pt$_{3}$V 
alloy at 1393 K a simple peak should be observed (see Eq.~(\ref{5})). 
Note, however, that these estimates are not fully consistent, since the 
interaction parameters for the two alloys are similar but not identical. 
Furthermore, in the approximate Eq.~(\ref{9a}) the experimental 
values of the SRO parameters were used, though they should have instead 
been calculated self-consistently. Nevertheless, Eq.~(\ref{8}) clearly 
shows the tendency for the self-energy curvature to become greater 
than the curvature of the interaction term when the concentration of 
Pt increases. 

\begin{figure}
\begin{center}
\includegraphics[angle=-90]{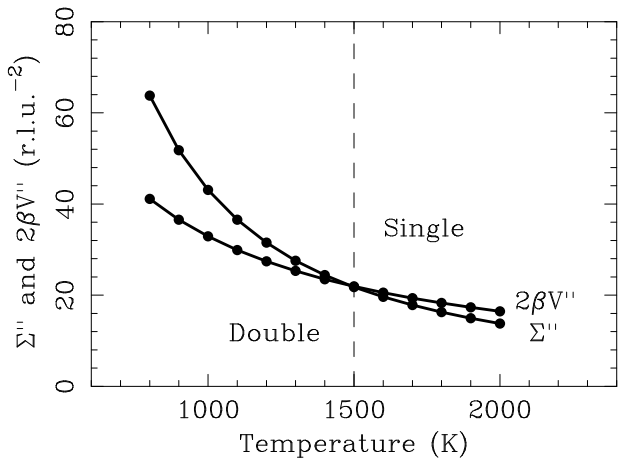}
\end{center}
\caption{The same derivatives as in Fig.~\ref{f1} but as functions 
of temperature for $c=1/9$.}
\label{f2}
\end{figure}
 
\begin{figure}
\begin{center}
\includegraphics[angle=-90]{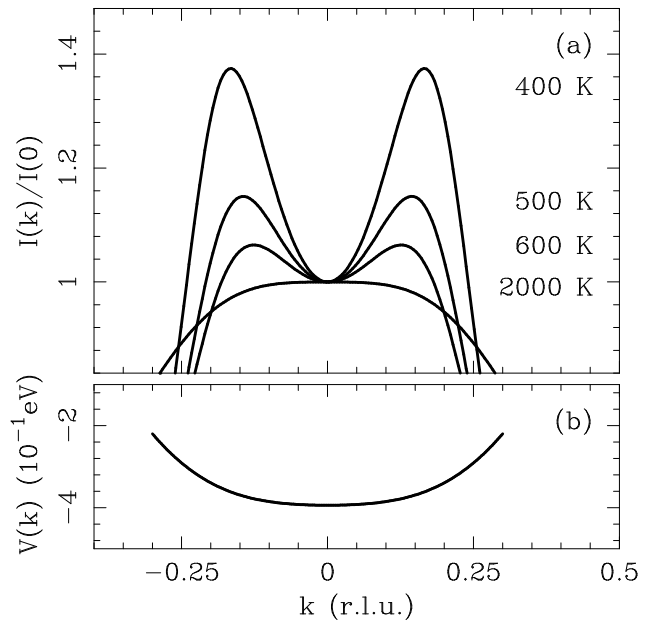}
\end{center}
\caption{(a) Normalized intensity $I(k)/I(0)$ for the Pt$_{8}$V 
composition at indicated temperatures and (b) the interaction 
$V(k)$ for the Pt$_{3}$V alloy.}
\label{f3}
\end{figure}
 
To confirm these preliminary considerations, we carry out the 
self-consistent AE calculations. The complete set of the AE equations 
constituted by Eqs.~(\ref{1}), (\ref{16}) and (\ref{9}) is solved for 
given interaction parameters, temperature and concentration to find 
the SRO parameters,  self-energy and intensity. A particular 
AE approximation is defined by neglecting fourth- and higher-order 
terms and including only finite number of coordination shells in 
Eqs.~(\ref{9}) for $\Sigma$. The zero-order approximation of the AE 
approach ($\Sigma_{lmn}=0, lmn \neq 000$) coincides with the 
SM~\cite{tokar1}. Here all calculations are performed using the 
set of interactions $V_{lmn}$ for the Pt$_{3}$V alloy (Table~\ref{t1}) 
and nine-shell approximation for the self-energy; inclusion of additional 
shells leads to negligible corrections.

\begin{figure}
\begin{center}
\includegraphics[angle=-90]{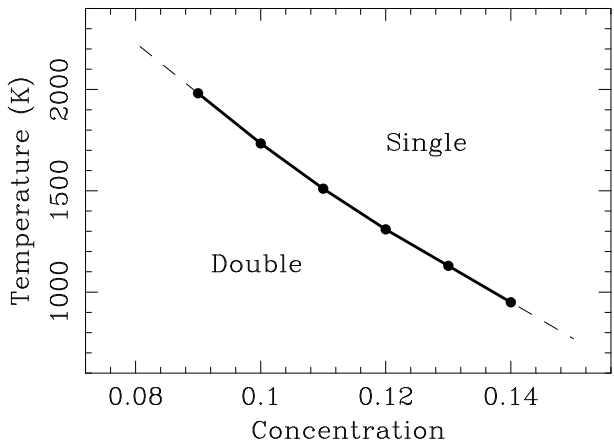}
\end{center}
\caption{``c-T phase diagram'' for the (100) intensity peak.}
\label{f4}
\end{figure}

The results of the calculations are shown in Figs.~\ref{f1}-\ref{f4}.
Fig.~\ref{f1} presents behaviour of the curvatures 
$\Sigma^{\prime \prime}$ and $2 \beta V^{\prime \prime}$ with 
concentration at 1224 K. The $\Sigma^{\prime \prime}(c)$ curve crosses 
the horisontal line $2 \beta V^{\prime \prime}=26.9$ r.l.u.$^{-2}$ (the 
derivative $2 \beta V^{\prime \prime}$ is concentration-independent)
at approximately $c=1/8$; according to Eq.~(\ref{5}), at this point the 
second derivative $I^{\prime \prime}$ vanishes. At greater concentrations 
$\Sigma^{\prime \prime} < 2 \beta V^{\prime \prime}$, so that 
$I^{\prime \prime} < 0$ and the intensity has a simple maximum 
at the (100) position. On the contrary, when the concentration is below 
this value, as in the case of the Pt$_{8}$V alloy, the curvature of the 
self-energy exceeds that of the interaction term, and the (100) 
intensity peak splits along the (h00) line. This is precisely what 
was found in Ref.~\cite{lebolloch}.

Furthermore, very simple arguments show that the splitting of the 
(100) intensity peak can develop also when temperature decreases at 
fixed concentration. Indeed, we have just seen that this peak splits 
for the Pt$_{8}$V composition at 1224 K if the interaction for the 
Pt$_{3}$V alloy which has a simple minimum at the (100) position is 
used. However, at high enough temperatures the asymptotically correct 
KCM formula~(\ref{3}) is valid, and the shape of the intensity follows 
that of $V({\bf k})$. This means that the transition from single- 
to double-peak structure of the (100) intensity maximum occurs with 
decreasing temperature. This conclusion is illustrated in Fig.~\ref{f2}, 
where the same curvatures as in Fig.~\ref{f1} are shown as functions 
of temperature for the Pt$_{8}$V composition. The compensation of the 
curvatures of the two terms in Eq.~(\ref{1}) occurs here at about 
$T=1500$ K. Profiles of the normalized intensity $I(k)/I(0)$ at 
different temperatures and of the interaction $V(k)$ are displayed in 
Fig.~\ref{f3}. The positions $\pm k_{0}$ of the intensity maxima and 
ratio $I(k_{0})/I(0)$ at 400 K agree well with the MC values $k_{0}=0.2$ 
r.l.u. and $I(k_{0})/I(0)=1.3$ for 410 K estimated from Fig.~4 in 
Ref.~\cite{lebolloch}. Fig.~\ref{f4} summarizes the findings of the 
present work, showing what can be called a ``c-T phase diagram'' for 
the (100) intensity peak. The line there is a set of points at which 
$I^{\prime \prime}=0$; crossing this line in any direction changes 
qualitatively the fine structure of the (100) maximum (single to double 
peak or vice versa).
 
\begin{figure}
\begin{center}
\includegraphics[angle=-90]{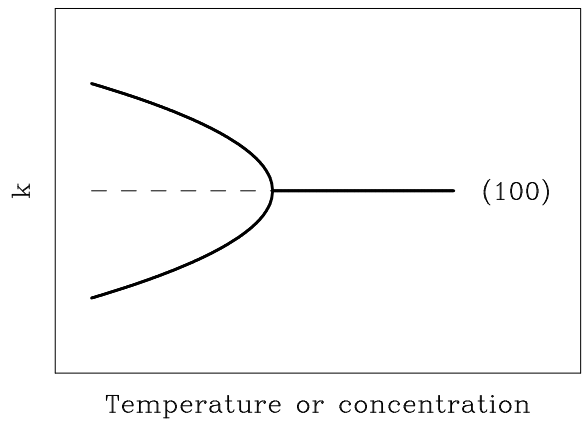}
\end{center}
\caption{ Schematic temperature (concentration) dependence 
of the intensity peak positions.}
\label{f5}
\end{figure}

It is necessary to note that the positions of the resulting $I({\bf k})$ 
peaks depend on the temperature and the concentration, as follows, 
e.g., from Fig.~\ref{f3}a. Here it is possible to draw a formal analogy 
with the Landau theory of second-order phase transitions~\cite{landau}. 
Let us consider the change of the intensity with temperature and assume that 
$I^{\prime \prime}$ vanishes at some temperature $T_{0}$. Then, for 
temperatures close to $T_{0}$ and wavevectors near the (100) position, 
we have
\begin{eqnarray}
I^{-1}(k) & = & I^{-1}(0) + \frac{1}{2} A k^{2} 
+ \frac{1}{4} B k^{4} \ , \label{14}
\end{eqnarray}
where $A=c(1-c)[-\Sigma^{\prime \prime}+2\beta V^{\prime \prime}]=
\tilde{A}(T-T_{0})$, $\tilde{A}$ and $B$ are positive constants, and 
only the lowest-order terms are retained in the expansions of $A$ and 
$B$ in powers of $T-T_{0}$. The same expansion can be written using 
concentration instead of temperature. The inverse intensity thus has 
exactly the form of the Landau free energy functional, with $k$ playing 
the role of the order parameter. Therefore, at small negative $T-T_{0}$ 
values the splitting increases as $(T_{0}-T)^{1/2}$. Contrary to the 
corresponding result of the genuine Landau theory, the obtained bifurcation 
exponent is exact, since the intensity is an analytical function of 
the wavevector and can legitimately be expanded into the Taylor series. 
The temperature (concentration) dependence of the peak positions is 
shown schematically in Fig.~\ref{f5}.

In conclusion, we have put forward an explanation of the experimental
and MC observations concerning the unusual concentration dependence 
of SRO diffuse scattering from the Pt-V alloy system. It was found 
in Ref.~\cite{lebolloch} that the same set of interaction parameters 
could produce qualitatively different intensity distributions at 
different concentrations, in this case simple and split (100) 
intensity peaks for the Pt$_{3}$V and Pt$_{8}$V compositions, 
respectively. In this Letter the observed anomaly is attributed to 
the competition between the reciprocal-space curvatures of the two 
terms in the expression for the SRO diffuse intensity. Currently used 
analytical theories of SRO neglect one of the curvatures, that of the 
self-energy (i.e. the ${\bf k}$-dependence of the latter); it has 
been shown how to overcome this difficulty by using the AE approach. 
The proposed theory predicts an analogous temperature dependence of 
the intensity and varying positions of the resulting intensity spots 
in the split-peak regime.

The author thanks D. Le Bolloc'h for communicating results of 
Ref.~\cite{lebolloch} prior to publication and useful correspondence.

\end{document}